\documentclass[conference]{IEEEtran}
\IEEEoverridecommandlockouts
\usepackage{cite}
\usepackage{amsmath,amssymb,amsfonts}
\usepackage{algorithmic}
\usepackage{graphicx}
\usepackage{textcomp}
\usepackage{xcolor}
\usepackage{rotating}
\usepackage{array}
\makeatletter
\newcommand{\thickhline}{%
 \noalign {\ifnum 0=`}\fi \hrule height 1pt
 \futurelet \reserved@a \@xhline
}
\newcolumntype{"}{@{\hskip\tabcolsep\vrule width 1pt\hskip\tabcolsep}}
\makeatother
\newcommand{\tabincell}[2]{\begin{tabular}{@{}#1@{}}#2\end{tabular}} 
\usepackage{amssymb} 

\def\BibTeX{{\rm B\kern-.05em{\sc i\kern-.025em b}\kern-.08em
 T\kern-.1667em\lower.7ex\hbox{E}\kern-.125emX}}
\begin{document}

\title{Security of Medical Cyber-physical Systems: \\An Empirical Study on Imaging Devices\\
\thanks{The authors would like to thank the vendors and developers for their help in the research. This research was financially supported by the National Key Research and Development Plan (2018YFB1004101), Key Lab of Information Network Security, Ministry of Public Security (C19614), Special fund on education and teaching reform of Besti (jy201805), the Fundamental Research Funds for the Central Universities(328201910), China Postdoctoral Science Foundation funded project, 2019 Beijing Common Construction Project-Teaching Reform and Innovation Project for Universities in Beijing, Key Laboratory of Network Assessment Technology of Institute of Information Engineering, Chinese Academy of Sciences.}
}

\author{\IEEEauthorblockN{Zhiqiang Wang$^{1,*,\dagger}$, Pingchuan Ma$^{1,*}$, Xiaoxiang Zou$^{2,\dagger}$, Tao Yang$^{3}$}

\IEEEauthorblockA{
 \textit{$^{1}$ Beijing Electronic Science and Technology Institute}\\
 \textit{$^{2}$ National Computer Network Emergency Response Technical Team/Coordination Center of China}\\
 \textit{$^{3}$ Key Lab of Information Network Security, Ministry of Public Security}\\
 \textit{$^{*}$ equal contributions}\\
 \textit{$^{\dagger}$ corresponding authors}
 }
}

\maketitle

\begin{abstract}
Recent years have witnessed a boom of connected medical devices, which brings security issues in the meantime. Medical imaging devices, an essential part of medical cyber-physical systems, play a vital role in modern hospitals and are often life-critical. However, security and privacy issues in these medical cyber-physical systems are sometimes ignored.

In this paper, we perform an empirical study on imaging devices to analyse the security of medical cyber-physical systems. To be precise, we design a threat model and propose prospective attack techniques for medical imaging devices. To tackle potential cyber threats, we introduce protection mechanisms, evaluate the effectiveness and efficiency of protection mechanisms as well as its interplay with attack techniques. To scoring security, we design a hierarchical system that provides actionable suggestions for imaging devices in different scenarios. We investigate 15 devices from 9 manufacturers to demonstrate empirical comprehension and real-world security issues.
\end{abstract}


\section{Introduction}

In the past decades, PACS/RIS (Picture Archiving and Communications System/Radiography Information System) has gone through digital evolutionary improvements, including richer function, better user experience and advances in security. However, few mechanisms were deployed to ensure system security and data privacy. Farhadi et al.~surveyed the security situation in several Iranian hospitals and stated that these medical information systems had the patient health data while lacking in suitable mechanisms to prevent security risks \cite{Farhadi2013}.

As these systems collect and manage highly sensitive data and contain a number of embedded software and hardware, here, we refer these systems to \emph{medical imaging cyber-physical systems (MICPS)} which have great demands on system security and data privacy as they are life-critical, context-aware and networked medical imaging devices \cite{lee2011challenges}.

DICOM\footnote{https://www.dicomstandard.org/} (Digital Imaging and Communications in Medicine), an international standard for imaging data transmission and storage, employed encryption as its only protection mechanism against cyber threats for more than 20 years, while it introduced more security mechanisms in its latest version. And in recent years, the industrial community begins to highlight security concerns of medical imaging devices. More standards were proposed to ensure the ability of imaging devices against attacks, e.g., IEC TR 80001-2-2:2012 and HIMSS/NEMA Standard HN 1-2013. However, the major concern of mentioned standards is about compliance rather than real-world security issues and whole lifecycle security. So, it is urgent to identify the way adversaries attack and the trade-off of defence against cyber threats for the medical imaging cyber-physical systems.

Prior to the work, the authors presented an exploratory overview of security issues of medical imaging systems and discovered some security issues of devices from well-known vendors \cite{wang2018medical}. This initial investigation showed that diagnostic imaging systems are on fire. However, the work didn't illustrate a comprehensive overview of threat models and protection mechanisms as well as detailed analysis.

\textbf{Contribution} The authors conveyed an empirical study of medical imaging devices to understand security in medical cyber-physical systems in this paper. In summary, the main contributions are:
\begin{itemize}
 \item we design a threat model and propose prospective attack techniques for medical imaging devices;
 \item we introduce protection mechanisms, evaluate the effectiveness and efficiency of protection mechanisms as well as its interplay with attack techniques;
 \item we design a hierarchical system that provides actionable suggestions for imaging devices in different scenarios;
 \item we investigate 15 devices from 9 manufacturers to demonstrate empirical comprehension and real-world security issues.
\end{itemize}

\section{Threat Model}

\begin{figure}[htbp]
 \begin{center}
 \includegraphics[width=\columnwidth]{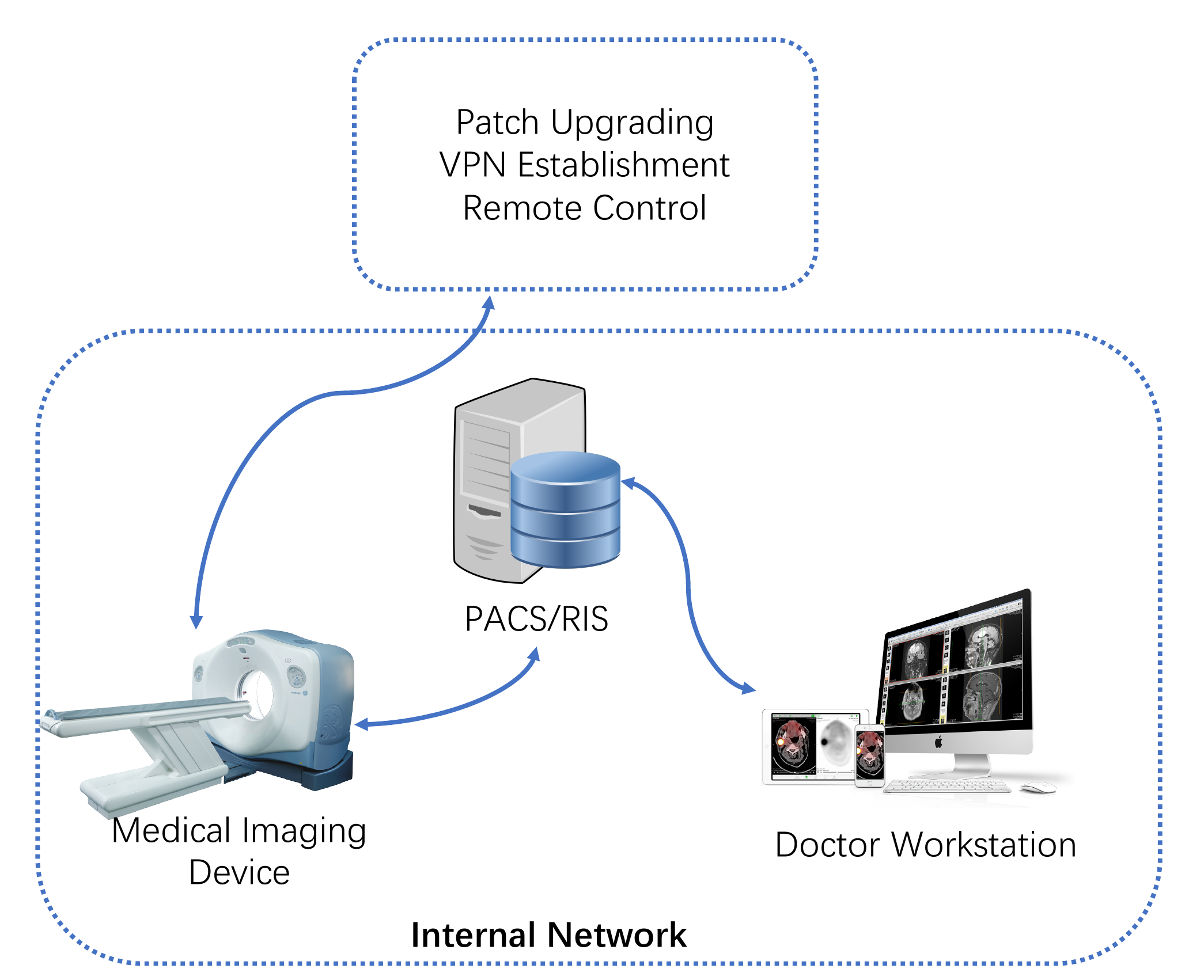}
 \caption{Visualization of a typical medical imaging device in a hospital network. Lines between devices/systems indicate data/control message transmission.}
 \label{fig:threatm}
 \end{center}
\end{figure}

Medical imaging devices are expected to be used in hospital internal networks and collaborate with systems inside and outside hospital networks, respectively. As is shown in Fig.~\ref{fig:threatm}, medical imaging devices are connected to hospitals' PACS/RIS and vendors' servers outside the hospital network for patch upgrading, VPN establishment and remote control. The PACS/RIS in hospital archives and renders the medical images, from which doctors' workstations access medical images.

\subsection{Attack Vector}

Based on the workflow, three types of attacks might be feasible.

\textbf{Remote servers poisoning.} In most cases, the devices are designed to receive upgrade patches, establish VPN (Virtual Private Network) with remote servers and execute remote commands. To this end, these systems maintain command and control interface to remote servers. Once vendors' remote servers are exploited by attackers, they are able to push malware, hijacking network traffic and execute remote control.

\textbf{Internal network penetration.} Many hospitals provide public WiFi to patients while don't employ network isolation strategy. So, attackers are possible to perform port scanning and social engineering to access these critical systems. In addition, ARP spoofing is possible in this scenario to perform MitM (man-in-the-middle) attack.

\textbf{Physical brute-force.} In some cases, malicious users are granted physical access permission to these systems. For example, although the attackers don't know the PIN code of a medical imaging device, he can view health data by taking the machines apart and getting the disk. Also, most devices enable the ``Emergency Access'' mode which provides basic functions without any authentication. Accordingly, sufficient physical defence mechanisms should be designed to avoid physical brute-force.

\subsection{Attack Techniques}
Here, we present three standard workflows to attack these devices within a network.

\subsubsection{Port Scanning}
Port scanning is the most popular technique for penetrating systems which probes a server or host for open ports. A large body of tools like Nmap\footnote{https://nmap.org} are developed and vary in rule-libraries for port scanning. With the help of well-designed tools, attackers can identify the service as well as its version on an open port and leverage its known vulnerabilities, e.g., OpenSSL HeartBleed Vulnerability, to exploit it.

\subsubsection{Traffic Analysis}
In practice, the medical imaging device is linked with an Ethernet line, and another Ethernet line from the mirror port of the switch is linked with the workstation to analyse all data transmission using Wireshark\footnote{https://www.wireshark.org}. However, some devices enable SSL communication, and attackers need to install private certification to decrypt encrypted data. Besides, some devices utilise VPN (Virtual Private Network) to send health data which makes traffic analysis not available in this case.

\subsubsection{Reverse Engineering}
Reverse engineering is the process of analyzing software to identify the interrelationships of different components and to discover security vulnerabilities. It is regarded as a practical approach to exploit and penetrate target devices. Usually, most software is presented in forms of binary code, and some are even obfuscated to prevent source code leakage. So, it is an extremely time-consuming task, especially when the system is vast and complicated. However, these systems are often installed with open-source software. Hence, attackers simply discover vulnerabilities in such software and attack target systems \cite{wang2019medical}.

\section{Protection Mechanisms}

\subsection{Encrypted Data Storage \& Transmission}

From the perspective of health data security, encryption methods are first applied to increase security during data transmission and storage. 
For example, manufacturer \emph{A} provides OpenVPN-enabled communication between medical imaging devices and PACS/RIS system, when the security mode is turned into the highest security level, so-called \emph{DoD Mode}. As a result, traffic is naturally encrypted. Some devices provide \emph{https}-based communication to enable secure transmission. 
In terms of storage encryption, the developers claimed that users could install the hardware security module (HSM) additionally, which enables data encryption feature; however, it is not installed by default.

\subsection{Network Protection}

Connected with a local area network, medical devices are likely to suffer from man-in-the-middle attacks. For medical imaging devices, a medical image standard, namely DICOM, enables node authentication in its file header. DICOM uses \emph{Application Entity Title} (\emph{AE Title}) to identify the DICOM nodes communicating between each other. To be precise, devices utilise \emph{AE Title}, an IP address and a port number to identify a certain node in the network.
Indeed, this mechanism cannot defence the ARP spoofing. Attackers can use ARP spoofing to create a fake node with a correct IP address and send DICOM file with accurate \emph{AE Title} at a correct port. Hardly can devices distinguish the malicious node unless other mechanisms are deployed. In other words, \emph{AE Title}-based mechanism only prevents the case that caused by errors rather than malicious attacks. More effective methods for medical imaging device node authentication remain to be studied.

\subsection{Physical Safeguards}

As is mentioned in the previous section, under the assumption that attackers can have physical access to the devices, it is necessary to deploy safeguards in case of brute-force attacks. 
Typically, devices are based on a workstation from some computer manufacturers such as hp. Hence, these workstations are pre-installed with a physical lock to prevent illegal access to hardware.

\subsection{System Hardening}

``System Hardening'' refers to a series of software-based protection mechanisms. Specifically, we investigate the function of the firewall, shortcut closure, patch installation and anti-virus applications. Shortcuts need to be closed because attackers can escape from the current interface and create a malicious process by some shortcuts, e.g., \emph{Ctrl+Shift+Delete} in Windows system calling TaskManager and starting processes.

\subsection{Security Guidance}

The reason we choose security guidance as a part of device protection mechanisms is that human plays a vital in security management and useful guidance can help users have a better understanding of their security situations and make the better configuration in their context. Current device guidance cares more about the functions of devices and only presents little security knowledge, and it is hard for a user without information security knowledge to config the extremely complex devices.

\section{Evaluation and Interplay between Attack Techniques and Protection Mechanisms}

\subsection{Evaluation}
\begin{figure}[htbp]
 \begin{center}
 \includegraphics[width=\columnwidth]{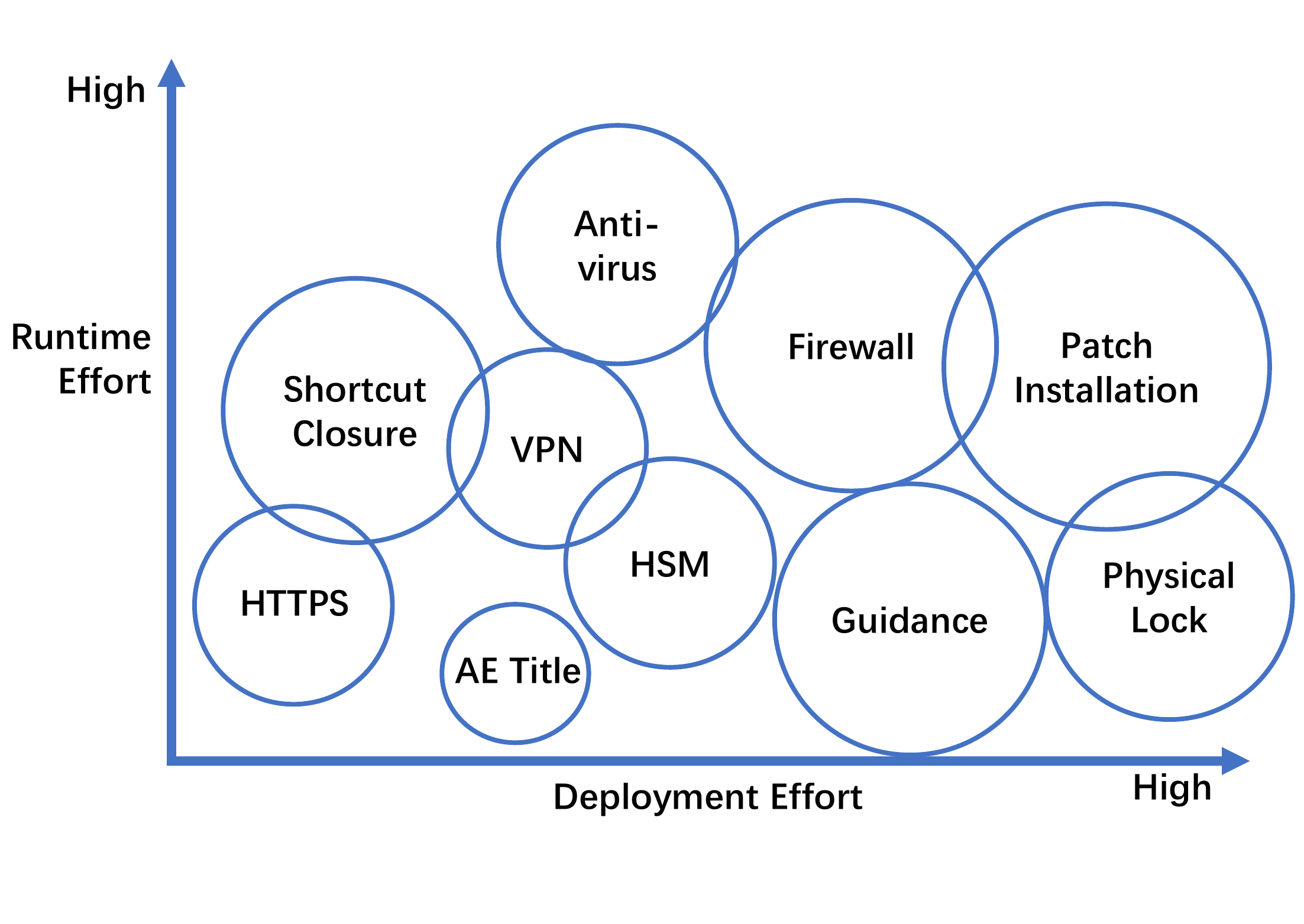}
 \caption{Evaluation of Protection Mechanisms}
 \label{fig:eva}
 \end{center}
\end{figure}

We refer to the evaluation framework proposed by Yuan et al. \cite{tian2016swords}. Their dimensions of evaluation contain deployment effort as well as runtime effort, and in this paper, we also consider the effectiveness of protection mechanisms. The cost of each protection mechanism includes the developer’s effort, the runtime cost and the effectiveness, which are the three dimensions shown in Fig.~\ref{fig:eva}. To be precise, the area of circles indicates the effectiveness of protection mechanisms. Based on our study, different mechanisms vary in places and circle sizes.

\textbf{Deployment effort} is a term that presents an important criterion to evaluate the performance of a protection mechanism. For developers, the primary objective is to build secure devices at as little as possible cost. The deployment cost guides developers to adopt mechanisms which are easy to implement. We would consider the criterion from the perspectives of the change occurred on original systems and the cost of equipment.

\textbf{Runtime effort} is a term that shows how much performance would be influenced by a specific mechanism. It is important to maintain excellent user experience and enable fast response of devices. Specifically, hardware-assisted mechanisms do not take into account of the runtime effort. Generally, we mainly estimate the runtime effort by the volume of performance loss.

\textbf{Effectiveness} is the most important criterion to present a comprehensive evaluation of protection mechanisms. Circles in different sizes represent the effectiveness of protection mechanisms. The detailed analysis is given previously.

\textbf{Analysis} \emph{https} is much easier to develop when compared with building a VPN server. Besides, VPN also requires more runtime efforts than the figure of \emph{https}. However, VPN slightly outperforms \emph{https} because it won't be attacked in the case that user installs a malicious certification though it is quite uncommon. In terms of HSM, the encryption together with decryption is completed by external hardware. Thus, the runtime efforts are lower than others' counterparts. 
Integrated into DICOM standard, \emph{AE Title}-based protection mechanism tends to take only a little runtime efforts and deployment efforts to authenticate remote node.
As a means of hardware-assisted mechanisms, similarly, the physical lock doesn't need many runtime efforts. But it is not easy to install if not pre-installed by workstation vendors.
Firewall and anti-virus play an essential role in device protection. It is not easy to develop a firewall or anti-virus software in the customized system environment and consumes computing resource to filter malicious traffic. But it is supposed to be taken into consideration when users' security need is high.
Patch installation takes a high number of deployment efforts as well as runtime efforts. Besides, some compatibility problems may take place when some software is upgraded. Despite the enormous efforts, it is still worth and necessary to be adopted for its great effectiveness.

\subsection{Interplay}

To have a deeper comprehension of the interplay between protection mechanisms and attack techniques, we present the detailed ability of protection mechanisms against cyber threats in Table.~\ref{tbl:inter}.

To begin with, VPN and \emph{https} almost enable the same feature in terms of secret communication, while VPN provides better resistance against insecure certification. HSM utilizes a hardware-assisted approach to prevent data from illegal access. The attack technique corresponded with \emph{AE Title} is hard to summarize in that it is more an error-correction mechanism than a protection mechanism. We argue that firewall can prevent port scanning as well as reverse engineering, because some firewall provides a malicious traffic filter, which, to some extent, contributes to the defence of reverse engineering. Similarly, patch installation also protects devices from being attacked by reverse engineering. Timely upgrading can fix vulnerabilities and provides a more secure system. Security guidance is not a part of the hardware and software systems of the devices. Nevertheless, from the perspective of products, appropriate security guidance plays a fundamental role to prevent threats introduced by human factors.

\begin{sidewaystable}[htbp]
 \centering
 \renewcommand\arraystretch{1.5}
 \caption{Interplay between Protection Mechanisms and Attack Techniques.}
 \label{tbl:inter}
 \setlength{\tabcolsep}{3pt}
 \begin{tabular}{c|c|c|c|c|c}
 ~& 
 Port Scanning&
 Traffic Analysis&
 Reverse Engineering&
 Physical Brute-Force&
 Others\\

 \thickhline

 VPN & & $\checkmark$ & & & Insecure Certification \\
 \hline
 \emph{https} & & $\checkmark$ & & & \\
 \hline
 HSM & & & & $\checkmark$&\\
 \hline
 \emph{AE Title} & & & & & \\
 \hline
 Physical Lock & & & &$\checkmark$ &\\
 \hline
 Firewall & $\checkmark$ & & $\checkmark$ & &\\
 \hline
 Anti-virus & & & & & Virus\\
 \hline
 Shortcut Closure & & & & & Interface Escape\\
 \hline
 Patch Installation & & &$\checkmark$ & & Vulnerability Exploitation\\
 \hline
 Security Guidance & & & & & Human Factors
 \end{tabular}
\end{sidewaystable}

\section{Hierarchical System}

\begin{table}[htbp]
 \centering
 \renewcommand\arraystretch{1.5}
 \caption{Hierarchy of Device Protection}
 \label{tbl:inve}
 \setlength{\tabcolsep}{3pt}
 \begin{tabular}{c|c|c|c}
 Requirement& 
 CL1&
 CL2 &
 CL3 \\

 \thickhline

 Storage Encryption & $\bigcirc$ & $\bigcirc$ & $\checkmark$ \\
 \hline
 Transmission Encryption & * & $\bigcirc$ & $\checkmark$ \\
 \hline
 Node Authentication & * & $\bigcirc$ & $\checkmark$ \\
 \hline
 Physical Safeguards & * & $\bigcirc$ & $\checkmark$ \\
 \hline
 \tabincell{c}{Role-based\\Access Control} & $\bigcirc$ & $\checkmark$ & $\checkmark$ \\
 \hline
 \tabincell{c}{Identity Authentication\\(If applicable)} & $\checkmark$ & $\checkmark$ & $\checkmark$ \\
 \hline
 Data Traceability & * & $\checkmark$ & $\checkmark$ \\
 \hline
 Auditing & $\bigcirc$ & $\bigcirc$ & $\checkmark$ \\
 \hline
 System Hardening & $\bigcirc$ & $\bigcirc$ & $\checkmark$ \\
 \hline
 Security Guidance & $\bigcirc$ & $\bigcirc$ & $\checkmark$ 

 \end{tabular}
\end{table}

We present a hierarchical system for medical imaging device security evaluation in Table.~\ref{tbl:inve}. Note that ``*'' means the mechanism is not required at this level. ``$\bigcirc$'' means at least one mechanism is taken to enable the corresponding feature, while the mechanism(s) may not be effective enough. ``$\checkmark$'' means at least one effective mechanism is taken to enable the corresponding feature.

Whether a mechanism should be taken involves many considerations. To establish the rating system, we consult with experts with industry, government and academic background respectively and refer to relevant standards and regulations. For example, we use the notion of ``Node Authentication'' instead of \emph{AE Title} and add the concepts of ``Access Control'', ``Identity Authentication'' as well as ``Data Traceability''. 

For storage encryption, ``$\bigcirc$'' means that either some of the health data, including demographic data in the file header, is not encrypted, or the health data is encrypted by a weak algorithm, such as DES. ``$\bigcirc$'' in transmission encryption means that the communication protocol is out of date (such as SSL 3.0 and lower versions). Specifically, both \emph{https} and VPN at correct versions are regarded as ``\checkmark''. For devices at CL1, node authentication is not needed. For CL3 devices, mechanisms such as \emph{AE Title} are not adequate to provide the highest-level security protection. Identity authentication is needed for all levels of devices if applicable. It is not applicable for doctors to complete identity authentication with bloodied hands when some devices are used in surgery. Data traceability is required for CL2 and CL3 devices. Usually, UID in DICOM standard can help to identify the source. Comprehensive auditing, system hardening and security guidance are necessary for all devices, while some small print may be different for each level.

\section{Investigation}
\subsection{Device Overview}
We visited some manufacturers and evaluate their devices in terms of security and privacy. The manufacturers are international companies, which contain the most market shares in the global medical imaging device market. To protect their commercial reputation, we keep the vendor name anonymous.

\subsection{Result}

We tested 15 devices on the mentioned protection mechanisms as well as other aspects, e.g., authentication, data archiving. We placed a selective result in Table.~\ref{tbl:resu} and full results will be available in extended version due to lack of space. Note that ``Partial'' in Storage Encryption means that the devices only encrypt part of health data; ``Partial'' in Transmission Encryption means that the devices only encrypt data in some special context; ``Partial'' in System Hardening and Security Guidance means that not all requirements in the term are satisfied.

\begin{sidewaystable}[htbp]
 \centering
 \renewcommand\arraystretch{1.5}
 \caption{Result of Medical Imaging Device Security. }
 \label{tbl:resu}
 \setlength{\tabcolsep}{3pt}
 \begin{tabular}{c|c|c|c|c|c|c}
 Manufacturer& 
 Type&
 \tabincell{c}{Storage\\Encryption}&
 \tabincell{c}{Transmission\\Encryption}&
 Physical Lock&
 System Hardening&
 Security Guidance\\
 \thickhline
 A & DSA & N/A & Partial & Full & Partial & N/A\\
 \hline
 A & Mini CT & Partial & Partial & Full & Partial & N/A\\
 \hline
 B & DIC & Partial & N/A & N/A & Partial & N/A\\
 \hline
 C & DR & N/A & N/A & N/A & Partial & N/A\\
 \hline
 D & CT & Partial & N/A & Full & Partial & Partial\\
 \hline
 E & DR & Partial & Partial & Full & Partial & Full\\
 \hline
 F & Mammography X-Ray & N/A & Partial & Full & Partial & Full\\
 \hline
 G & Dental X-Ray & N/A & Partial & Full & Partial & Full\\
 \hline
 H & DR & Full & Full & N/A & Full & Full
 \end{tabular}
\end{sidewaystable}

As is shown in Table.~\ref{tbl:resu}, none of the devices can fully satisfy our requirements. Some devices only adopt fundamental protection mechanisms which can be exploited by some attack vectors. In terms of encryption, the fact is that only a small part of data is encrypted or well-protected. In terms of transmission, there is an essential point that while some devices design a grading system and adopt full support to the security requirements at the highest level, the default configuration only meets the lowest level where molecular mechanisms are adopted.

\subsection{Additional Security Issues}
Some critical issues in current systems are not presented in the table, and we describe as follows.

\begin{itemize}
 \item The implement of data restore is vulnerable as well. Most back-up data is stored by plaintext, and little validation mechanisms are considered.
 \item Auditing function is not well designed in many devices. Many devices even don't record user logging, and some records are duplicate or incomplete.
 \item Most devices pre-installed open-source software, such as OpenSSH, FTP Server and Samba. However, lacking in timely upgrading, these software suffers from some vulnerabilities (such as CVE-2016-10010\footnote{https://nvd.nist.gov/vuln/detail/CVE-2016-10010}, CVE-2012-0002\footnote{https://nvd.nist.gov/vuln/detail/CVE-2012-0002} and CVE-2017-0143\footnote{https://nvd.nist.gov/vuln/detail/CVE-2017-0143}).
\end{itemize}

\section{Related Work}
Fu et al.~firstly found that software radio can be applied to attack implantable cardioverter defibrillators (ICD) and proposed three approaches for defence. They reverse-engineered the ICD's communication protocol. Then software-defined radio attacks were applied to get patients’ medical information and even turn off devices. Various aspects of implantable medical device security have been further studied \cite{denning2008absence,burleson2012design,rushanan2014sok,denning2010patients,PMID:20357279}.

Sametinger et al.~gave a comprehensive overview of security challenges for medical devices \cite{sametinger2015security}. They presented challenges along with illustrative examples. However, the cases related to medical imaging devices are not described in detail as well, and system security implications of medical devices are not given. Kramer et al.~studied the security quality of medical devices under the administration of FDA Postmarket Guideline \cite{kramer2012security}. They mentioned a recall of a radiation therapy system in that “the product has a software problem in which previous patient measurement data gets associated with another patient’s image”.

Besides implantable medical devices, Tamara et al.~presented an experimental study of the surgical teleoperated robotic systems \cite{bonaci2015experimental}. They presented some Denial-of-Service attacks on the \emph{Raven II} robot. Then, Alemzadeh et al.~proposed a model-based analysis framework to detect and mitigate attacks \cite{7579758}.

Jagannathan et al.~presented an overview of generic medical device security frameworks and proposed a methodology called ``Cybersecurity Preliminary Hazards Analysis'' for medical devices cybersecurity assessment \cite{jagannathan2015cybersecurity}. Their method is based on 
guidelines and standards issued by official organizations and used for embedded devices in their paper.

\section{Conclusion}
In this paper, we test 15 representative devices and present a comprehensive study of medical imaging device security. We also compare the protection mechanisms implemented in current products and possible hacking techniques and propose the threat model to have a more comprehensive understanding. A hierarchical system is introduced for different scenarios. The system provides an actionable way for developers to improve their products and for government agencies to evaluate devices. The work we conveyed in this paper will be adapted in the industry as a part of pre-market security assessment for medical imaging products.

\bibliographystyle{plain}
\bibliography{conference_101719}

\end{document}